\documentclass{tMOP2e}
\usepackage{epsfig}
\usepackage{bm}
\begin{document}

\title{Ultra-cold fermions in optical lattices}

\author{P. D. Drummond, J. F. Corney, X.-J. Liu, H. Hu \\
 ARC Centre of Excellence for Quantum-Atom Optics, \\
 School of Physical Sciences, University of Queensland, \\
 Brisbane, QLD 4072, Australia}

\maketitle
\begin{abstract}
We summarize recent theoretical results for the signatures of strongly
correlated ultra-cold fermions in optical lattices. In particular,
we focus on: collective mode calculations, where a sharp decrease
in collective mode frequency is predicted at the onset of the Mott
metal-insulator transition; and correlation functions at finite temperature,
where we employ a new exact technique that applies the stochastic
gauge technique with a Gaussian operator basis. 
\end{abstract}

\section{Introduction}

Recent progress in obtaining bosonic and fermionic quantum gases has
ushered in a new experimental paradigm in physics. An essential feature
in these developments is that, although technically challenging, ultra-cold
atom experiments have remarkably simple physical conditions. The significant
issues are that:

\begin{itemize}
\item the underlying interactions are well understood, 
\item there are few parameters needed 
\item interactions can be tuned in many cases 
\end{itemize}
This will greatly help our understanding of many-body physics, since
precise theoretical models can be readily obtained at low density,
without the complications of a crystal lattice and reservoirs -- as
is so often the case at present.

Thus, one can apply simple theoretical models with high accuracy,
allowing innovative experimental tests of methods in quantum field
theory (QFT). This could lead to novel fundamental experiments in
physics, such as tests of massive particle quantum entanglement, as
well as applications to metrology.

An important element in recent investigations is the technique of
coherent transformation of cold atoms to molecules, either using photo-association\cite{Wynar}
or Feshbach\cite{Donley} resonance methods. This allows a strong,
tunable interaction between the atoms. Experiments have confirmed
coherently coupled atom-molecular models\cite{DKH98}, with renormalization\cite{Hollandetal}.

Bosonic molecular experiments include $^{133}$Cs, $^{87}$Rb, and
$^{23}$Na \cite{Bosonic-Cs-Rb-Na}. Fermi gases of $^{40}$K and
$^{6}$Li atoms have proved to be ideal experimental systems owing
to their long molecular lifetimes\cite{Petrov2004}. These have resulted
in cold molecule production\cite{Fermi-Mol}, molecular Bose-Einstein
condensation\cite{Fermionic-MolBEC}, and preliminary indications
of fermion superfluid behaviour \cite{MolBEC,BCS}.

The properties of ultracold Bose gases in optical lattices\cite{Jaksch1998}
are another `hot' topic, as a starting point for the exploration of
strongly correlated many-body systems\cite{greiner2002,Recati}. Fermionic
experiments in optical lattices are already underway\cite{Pezze2004},
thus allowing the direct realization of the strongly coupled lattice
of fermions known as the Hubbard model. In this paper we review some
recently obtained results for the Hubbard model, in particular:

\begin{itemize}
\item the use of collective modes as signatures for the metal-insulator
phase-transition 
\item a new exact Gaussian technique for calculating finite temperature
correlation functions, without the Fermi sign problem of traditional
Monte-Carlo methods. 
\end{itemize}

\section{Hubbard model}

A one-dimensional, sinusoidal optical potential is easily created
for trapped atoms. The results are different for fermions and for
bosons, since because of the Pauli exclusion principle, at least two
distinct spin eigenstates are needed to have on-site interactions
with fermions.

This leads to the (fermionic) Hubbard model,

\begin{equation}
\mathcal{H}=-\sum\limits _{j,\sigma=\pm1}t_{ij}\left(\widehat{a}_{i,\sigma}^{\dagger}\widehat{a}_{j,\sigma}+h.c.\right)+U\sum_{j}\widehat{n}_{j,+}\widehat{n}_{j,-}+\sum_{j,\sigma=\pm1}V_{j}\widehat{n}_{j,\sigma}\,\,,\end{equation}
 where $V_{j}$ includes the external trap potential which we assume
is harmonic, $U$ gives rise to on-site interactions, and $t_{ij}$
describes tunneling between the sites. This model of interacting fermions
is also used in high-$T_{c}$ superconductivity, where it is more
qualitative than quantitative in its applicability. The dimensionality
and lattice structure of the model determine the tunneling matrix
$t_{ij}$. In most of the paper, we will assume this is a one-dimensional
lattice with $t_{ij}=t\delta_{i+1,j}$, and $V_{j}=m\omega_{0}^{2}d^{2}j^{2}/2$,
where $m$ is the fermion mass, $\omega_{0}$ the trap frequency,
and $d$ the well-separation. We also give correlation results for
two-dimensional lattices in the last section.

\subsection{Band structure}

In a single-band model as given above, one can expect a band insulator
to form in the non-interacting limit, when all available sites have
been occupied (i.e., two particles per site). However, this is complicated
by the existence of the external potential. For a large, deep trap,
the density increases towards the centre, so that the insulating region
becomes localised, coexisting with a conducting region in the wings.

\begin{figure}
\begin{center}\centerline{\includegraphics[%
  width=6cm]{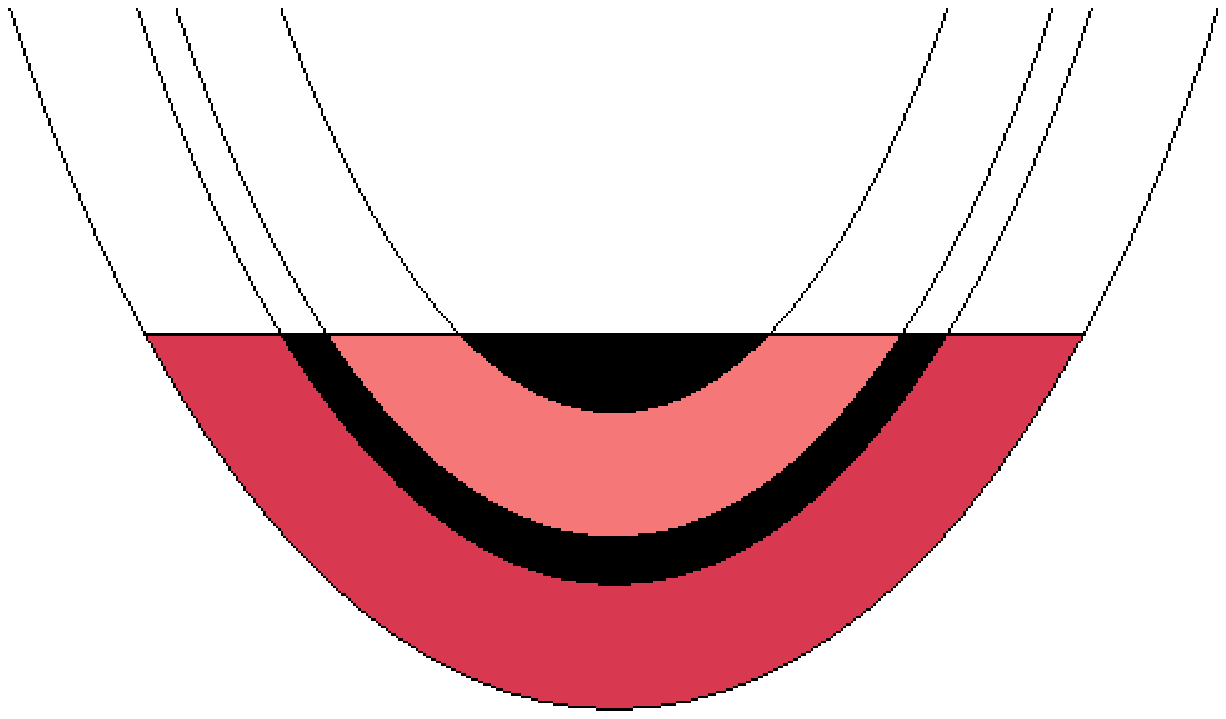}\includegraphics[%
  width=6cm]{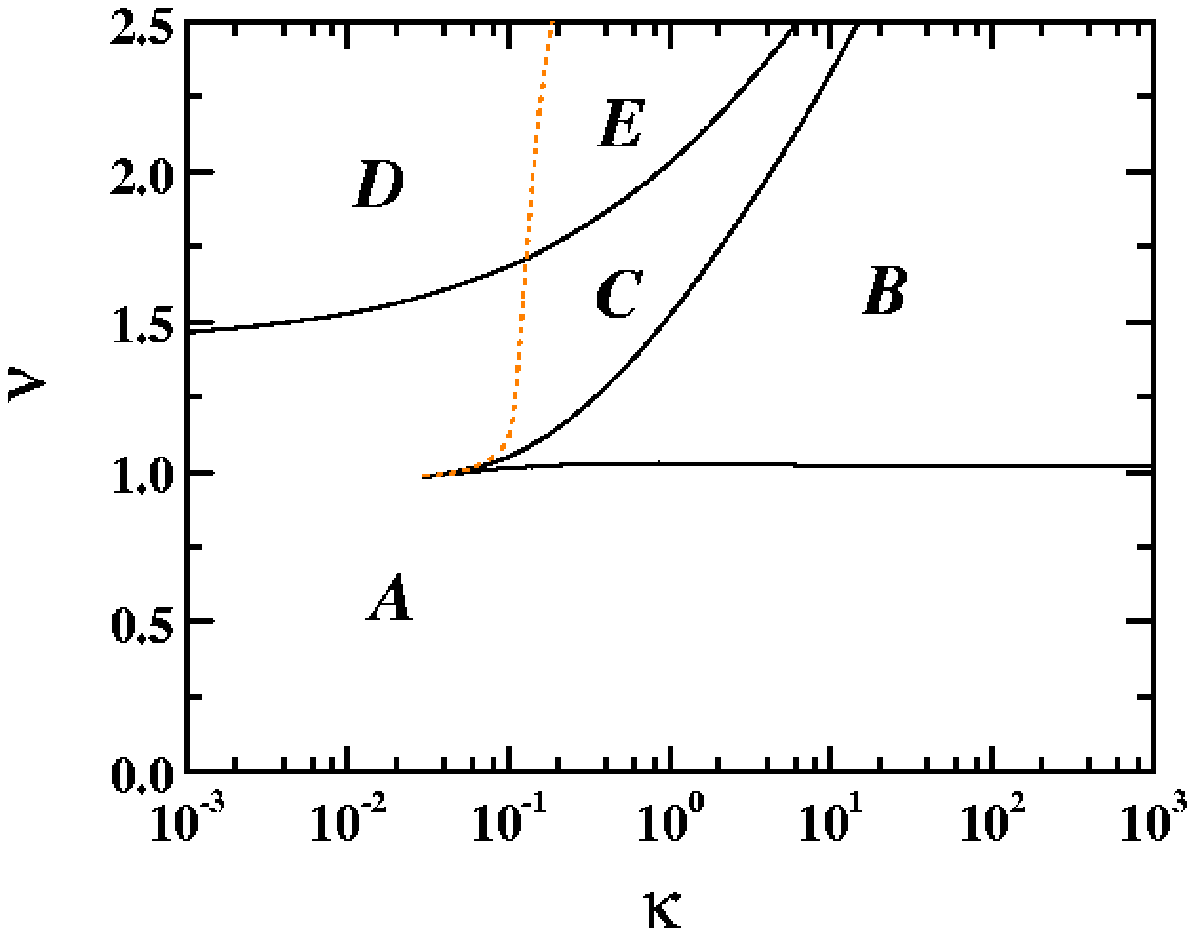}}\end{center}

\caption{Left graph: schematic illustration of effective single-particle band
structure in a harmonically trapped fermionic Hubbard model. Right
graph: phase-diagram vs the coupling ($\kappa$ ) and filling ($\nu$
). For details see Section (\ref{sub:Local-Density-Approximation}).\label{cap:Schematic-illustration-of}}
\end{figure}

When there are strong repulsive on-site interactions, the system develops
a new band structure in which an insulating region forms at half-filling,
i.e., with only one particle per site. This is shown in Figure (\ref{cap:Schematic-illustration-of}).

\subsection{Local Density Approximation}

\label{sub:Local-Density-Approximation}In the local density approximation
(LDA), we assume that the trap is slowly varying compared to all correlation
length scales. This allows us to use the piecewise exact solution
for the uniform 1D Hubbard model\cite{LiebWu68} to describe the strongly-correlated
interacting ground state.

It most useful to introduce an effective mass and frequency, corrected
for the band-dispersion, as well as a dimensionless coupling constant
and filling factor, where:

\begin{itemize}
\item Effective mass, frequency: $\, m^{*}=\hbar^{2}/\left(2td^{2}\right)$,
$\,\omega=\hbar\omega_{0}\left(m/m^{*}\right)^{1/2}/t$
\item Coupling constant $\,\kappa=U^{2}/\left(8t^{2}N\omega\right)$
\item Effective filling factor $\,\nu=\sqrt{2N\omega}/\pi$
\end{itemize}
The effective filling factor can be understood as corresponding to
a single particle per site ($n=1$) at the trap centre for $\nu=1$,
although the interpretation is more complex in general, due to the
effects of the trap potential. Note that $\nu$ is defined as a global
parameter for the entire trap, while the occupation per site ($n$)
varies with radius. 

With these parameters, and using the LDA, with a fixed global chemical
potential and harmonic trap potential, one obtains a phase-diagram\cite{Rigol,Xiaji}
as shown in the right-hand diagram in Fig (\ref{cap:Schematic-illustration-of}).
The labelled regions are as follows:

$A$: a pure metallic phase. In this case there are no filled bands,
and the quantum gas behaves analogously to a metallic conductor, with
$n<2$ everywhere.

$B$: a single Mott insulator domain at the center, accompanied by
two metallic wings. Here the strong particle-particle interactions
prevent occupation numbers larger than one, even though these are
allowed by the Pauli principle. 

$C$: a metallic phase at the center surrounded with Mott insulator
plateaus. In this case, a larger particle number opens up a second
band in the trap center, in which higher occupations with $n>1$ are
possible.

$D$: a band insulator at the center with metallic wings outside.
This is similar to phase B. Here, however, due to weaker inter-particle
interactions there is no Mott phase, and the central band insulator
with $n=2$ is caused purely by the Pauli exclusion principle.

$E$: a band insulator at the trap center surrounded by metallic regions,
in turn surrounded by Mott insulators. In this case there is a large
particle number \emph{and} strong interactions, so both Mott insulator
and band insulator regions can occur at different radii from the centre.

\section{Collective mode frequencies}

\subsection{Luttinger approximation}

In order to describe the collective modes, we can make use of the
Luttinger long-wavelength Hamiltonian\cite{Recati}, valid for small
displacements. Because of the spin-density separation in the Hamiltonian,
there are two types of waves present, described by the indices $\nu=\rho,\sigma$.
We assume here that the external potential only depends on the density,
and does not transform one type of wave into the other, so that the
inhomogeneous Hamiltonian is:\begin{equation}
{\mathcal{H}}_{\textrm{LL}}=\sum_{\nu=\rho,\sigma}\int dx\frac{u_{\nu}(x)}{2}\left[K_{\nu}(x)\Pi_{\nu}^{2}+\frac{1}{K_{\nu}(x)}\left(\frac{\partial\phi_{\nu}}{\partial x}\right)^{2}\right]\,\,.\end{equation}

Here the density and spin-wave velocities are $u_{\rho}$, $u_{\sigma}$respectively,
and these have Luttinger exponents $K_{\rho}$, $K_{\sigma}$. Using
the LDA, we can therefore solve the resulting wave-equations for the
collective mode frequency, with zero-current boundary conditions at
the boundaries where the density vanishes\cite{Combescot}.

\begin{figure}
\begin{center}\includegraphics[%
  width=12cm]{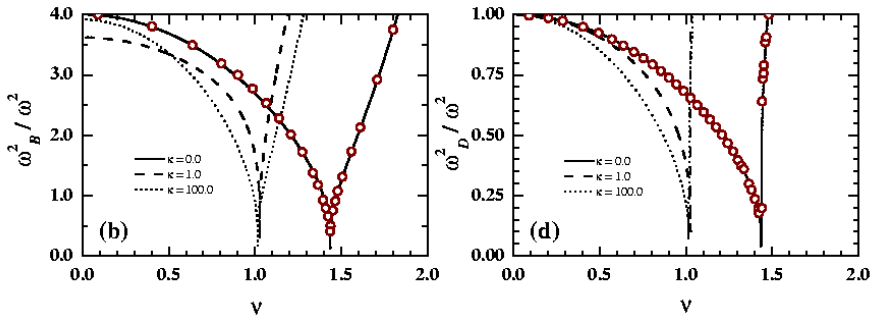}\end{center}

\caption{Breathing ($\omega_{B}$) and dipole ($\omega_{D}$) mode frequency
vs filling factor $\nu$ for zero, intermediate and strong couplings.
\label{cap:Collective-mode-frequency}}
\end{figure}

The results are shown in Fig (\ref{cap:Collective-mode-frequency}),
in which we have solved for the collective fermionic modes in a trapped
one-dimensional two-component Fermi gas, together with an optical
lattice\cite{Xiaji}. These graphs show a sharp frequency dip as an
unmistakable signature of Mott metal-insulator transition (MMIT) physics.
This occurs just at the filling factor where the insulating region
starts to form at the trap centre.

The reason for this is simple to understand. Collective modes involve
large density currents, which propagate through the lattice with a
characteristic velocity $u_{\rho}$. However, these velocities tend
to zero in the neighborhood of an insulating region, and must be exactly
zero inside the insulator. This effect slows down transport processes
and therefore reduces all collective mode frequencies, which scale
as $velocity/length$. The effect is strongest just when the insulating
region starts to form, leading to a large fraction of the conducting
region having a low average velocity. As the insulating region grows,
the part of the trap that is conducting is localized to the wings,
so the frequency increases again due to the smaller characteristic
lengths involved.

However, there are some limitations to these methods. In particular,
this is a linearized method, and hence is only valid for small displacements;
it also only applies to zero temperature.

Preliminary results in the non-interacting case using the Boltzmann
equations (circles in Fig (\ref{cap:Collective-mode-frequency}))
show that there are additional damping effects that are proportional
to displacement. This leaves an unsolved problem, of how to deal with
finite temperatures and large displacements in the interacting case.

\section{Quantum simulation with Gaussian operators}

We now turn to the question of how to carry out many-body quantum
field theory calculations without approximations. Quantum Monte-Carlo
(QMC) techniques for fermionic problems are strongly limited by the
Fermi sign problem, which causes large sampling errors\cite{FermiSignProblem}.

Here we give some results using a novel method constructed from a
basis of Gaussian operators, which treats \textbf{covariances} (Green's
functions) as phase-space variables. In principle, this method can
simulate both fermions and bosons\cite{bosefermi}, including either
thermal ensembles and dynamics. We will treat the case of finite-temperature
fermionic ensembles for definiteness.

\subsection{Gaussian expansion}

Our approach is different from traditional QMC methods. We expand
the state density operator $\,\widehat{\rho}$ in an operator basis
$\,\widehat{\Lambda}$:

\begin{equation}
\,\widehat{\rho}=\int P(\overrightarrow{\lambda})\widehat{\Lambda}(\overrightarrow{\lambda})d\overrightarrow{\lambda}\,\,,\end{equation}
 where $P(\overrightarrow{\lambda})$ is a probability distribution,
sampled stochastically over the variables$\overrightarrow{\lambda}$
which constitute a phase-space. Time (or inverse temperature) evolution
is obtained by a mapping procedure which maps operator equations into
stochastic equations:\begin{eqnarray}
\partial\widehat{\rho}/\partial t=\widehat{L}[\widehat{\rho}] & \longrightarrow & \partial\overrightarrow{\lambda}/\partial t=\overrightarrow{A}+\mathbf{B}\,\overrightarrow{\zeta}\,\,.\end{eqnarray}

These equations are typically not unique, and have a `stochastic gauge'
degree of freedom, which must be chosen to make the distribution compact
enough to eliminate boundary terms and minimize sampling errors\cite{gauge}.

\subsection{Gaussian operators}

The mapping procedure depends on the operator basis chosen, and we
shall choose a Gaussian operator basis, defined for both Fermi\cite{fermi}
and Bose\cite{bose} cases, as: \begin{equation}
\,\widehat{\Lambda}_{\pm}(\mathbf{n})=\Omega\left|\mathbf{I}\pm\mathbf{n}\right|^{\mp1}:\exp\left[\widehat{\mathbf{a}}\left(\mathbf{I}\mp\mathbf{I}-\left[\mathbf{I}\pm\mathbf{n}\right]^{-1}\right)\widehat{\mathbf{a}}^{\dagger}\right]:\,\,.\end{equation}

Here, annihilation~and~creation operators are included as a vector:
$\,\widehat{\mathbf{a}}=\left(\widehat{a}_{1},...,\widehat{a}_{M}\right)^{T}$.
The upper signs apply for bosons, the lower signs for fermions.

A typical application of these methods is to calculate thermal equilibrium
correlation functions. Since the fermion number is usually unknown,
the grand canonical distribution, $\widehat{\rho}=\exp(-(\widehat{H}-\mu\widehat{N})\tau)$,
is the most appropriate choice of ensemble. Here, $\widehat{\rho}$
is the unnormalised density operator, $\tau=1/k_{B}T$ is the inverse
temperature, and $\mu$ is the chemical potential. The phase-space
variable $\mathbf{n}$ is a complex matrix, which is best thought
of as a (finite temperature) stochastic Green's function. Thus, $\left\langle \,\widehat{a}_{i}^{\dagger}\widehat{a}_{j}\right\rangle _{\tau}=\left\langle \Omega n_{ij}\right\rangle _{P(\tau)}$,
where $\left\langle .\right\rangle _{P}$ indicates a phase-space
average over the distribution $P$.

\section{Fermionic correlations}

Rewriting the canonical density operator as an equation for inverse
temperature evolution, one obtains:\begin{equation}
\frac{d\widehat{\rho}}{d\tau}=-\frac{1}{2}\left[(\widehat{H}-\mu\widehat{N}),\widehat{\rho}\right]_{+}\,\,.\end{equation}
 After applying the appropriate identities, one then obtains the following
Stratonovich stochastic gauge equations for the Hubbard model, where
$\Omega$ is a weight and $\mathbf{n}_{\sigma}$ is the stochastic
Green's function for spin $\sigma=\pm1$:\begin{eqnarray}
\frac{d\Omega}{d\tau} & = & -\Omega H(\mathbf{n}_{+},\mathbf{n}_{-})\nonumber \\
\frac{d\mathbf{n}_{\sigma}}{d\tau} & = & \frac{1}{2}\left\{ \left(\mathbf{I}-\mathbf{n}_{\sigma}\right)\bm T_{\sigma}^{(1)}\!\mathbf{n}_{\sigma}+\mathbf{n}_{\sigma}\bm T_{\sigma}^{(2)}\!\left(\mathbf{I}-\mathbf{n}_{\sigma}\right)\right\} \,\,.\end{eqnarray}

Here we have introduced a transition matrix $T_{\sigma}^{(j)}$ and
real Gaussian noise terms defined by:

\begin{itemize}
\item T-matrix: $\, T_{i,j,\sigma}^{(r)}=t_{ij}-\delta_{i,j}\left\{ |U|(sn_{j,j,-\sigma}-n_{j,j,\sigma}+\frac{1}{2})-\mu+\sigma^{(s+1)/2}\xi_{j}^{(r)}\right\} ,$
\item Noises: $\,\left\langle \xi_{j}^{(r)}(\tau)\,\xi_{j'}^{(r')}(\tau')\right\rangle =2|U|\delta(\tau-\tau')\delta_{j,j'}\delta_{r,r'}\,\,$. 
\end{itemize}

\subsection{Finite-temperature correlations}

\begin{figure}
\begin{center}\includegraphics[%
  width=6cm,
  keepaspectratio]{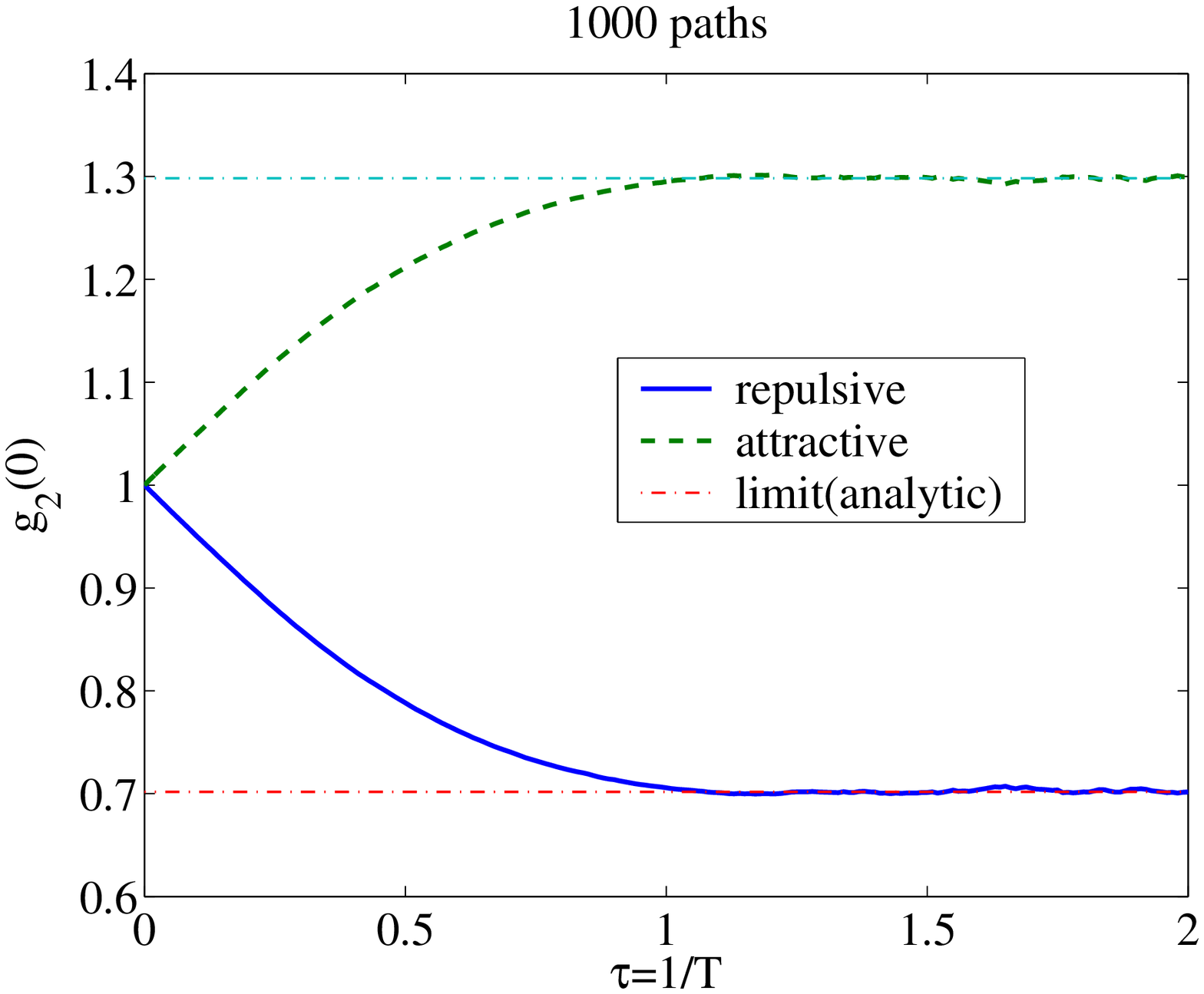}\includegraphics[%
  width=6cm,
  keepaspectratio]{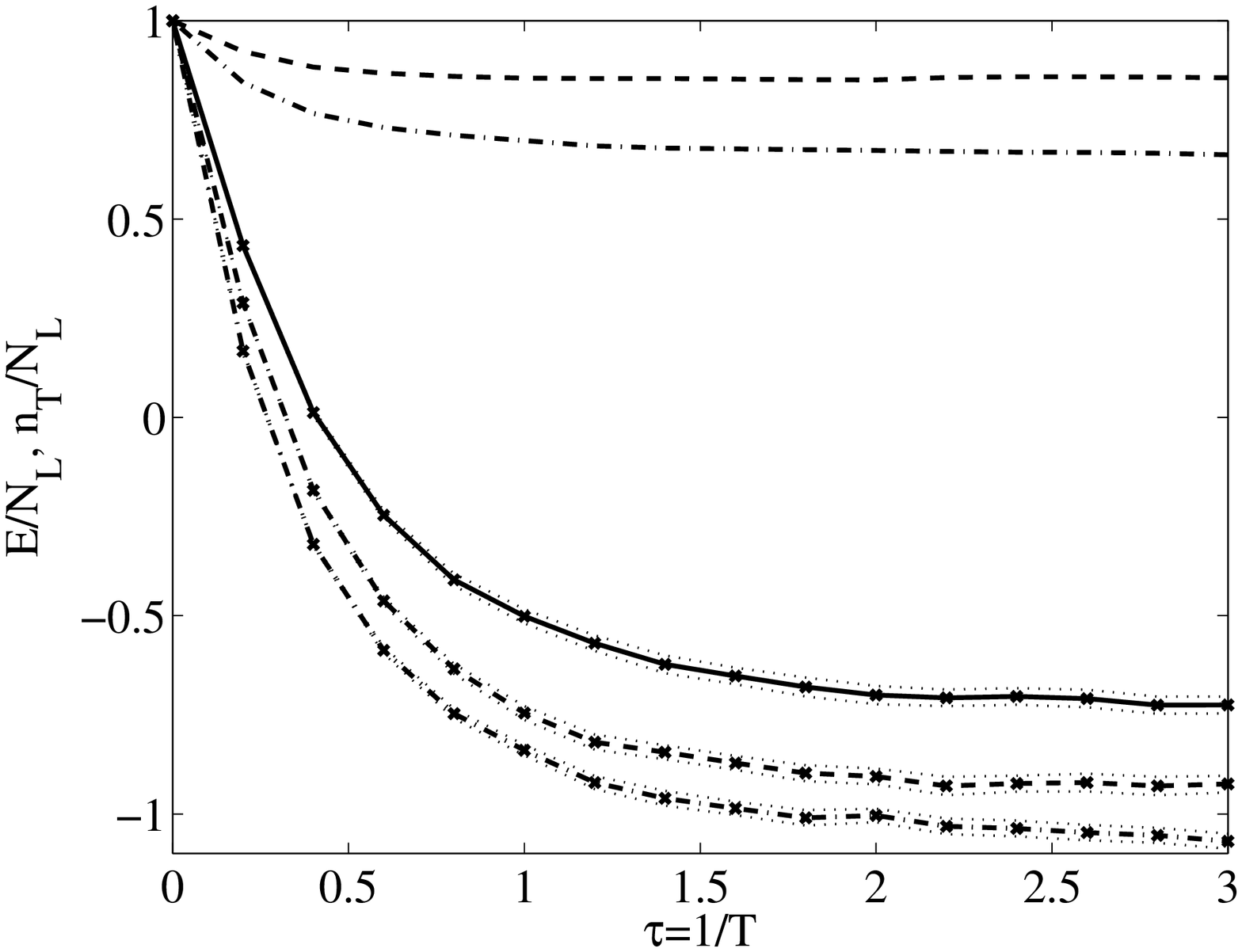}\end{center}

\caption{Hubbard models vs inverse temperature $\tau$. Left graph shows the
second-order correlation function $g_{2}(0)=\left\langle \widehat{n}_{jj+}\widehat{n}_{jj-}\right\rangle /\left\langle \widehat{n}_{jj+}\right\rangle \left\langle \widehat{n}_{jj-}\right\rangle $
for a $100$-site 1D lattice at half-filling. $|U|=2$ and $t=1$.
The right graph shows a $16\times16$ 2D lattice for chemical potentials
$\mu=2$ (solid), $\mu=1$ (dashed) and $\mu=0$ (dot-dashed). Curves
without crosses give the number of particles per site for $\mu=1$
(dashed) and $\mu=0$ (dot-dashed). $U=4$, $t=1$, and 50 paths initially.
\label{cap:Correlations-in-the}}
\end{figure}

The left graph of Fig (\ref{cap:Correlations-in-the}) shows the second-order
correlation function $g^{(2)}$ for a simulation of the uniform one-dimensional
Hubbard model, with both attractive and repulsive on-site interactions,
compared to the analytic result at zero temperature\cite{LiebWu68}.

The right graph shows the mean energies and particle numbers in a
$2D$ lattice for a variety of chemical potentials, with no evidence
for a Fermi sign problem that would cause enhanced sampling errors
at low temperature. We note that this extension to higher dimensions
is especially troublesome for traditional QMC techniques in the repulsive
case. Our technique shows no evidence of increased sampling error
at increased dimensionality, or with changes either to the filling
factor or the sign of the coupling constant.

\section{Summary}

We have reviewed some recent theoretical developments for ultra-cold
fermionic atoms in an optical lattice.

For the case of the Mott metal-insulator (MMIT) transition in one
dimension, at zero temperature, we give a universal phase diagram
based on the exact 1D solutions, and valid in the LDA limit of large
traps. A crucial experimental signature was predicted, namely a strong
frequency dip at the filling factor corresponding to the onset of
the MMIT at the trap centre.

More generally, a novel exact technique for general dynamic and static
Fermi calculations was presented. This can be used to calculate correlations
of strongly-correlated fermions at any temperature - and for Hubbard
models in any dimension. This is therefore a solution to the long-standing
Fermi sign problem in the Hubbard model.

\end{document}